\documentstyle[aps,prb,multicol,graphics]{revtex}

\begin{document}
\title{Buckling and d--Wave Pairing in HiTc--Superconductors}
\author{O. Jepsen, O.K. Andersen, I. Dasgupta, and S. Savrasov}
\address{Max-Planck Institut f\"ur Festk\"orperforschung, D--70569
Stuttgart, Germany}
\maketitle
\begin{abstract}
We have investigated whether the electron-phonon interaction can
support a d--wave gap--anisotropy. On the basis of models derived
from LDA calculations, as well as LDA linear-response calculations
we argue that this is the case, for materials with buckled or
dimpled CuO$_2$ planes, for the so--called buckling modes, which
involve out--of--plane movements of the
plane oxygens.
\end{abstract}

\begin{multicols}{2}

The mechanism of high-temperature superconductivity in hole-doped materials
containing CuO$_2$ planes remains a subject of vivid debate.
\cite{Plakida} 
A large amount
of experimental data such as superconductivity-induced phonon
renormalizations\cite{Cardona}, 
a large isotope effect away from optimal doping\cite{Frank}, and
phonon-related features in the tunnelling spectra\cite{Vedeneev} 
show effects of the
electron-phonon interaction. Density-functional (LDA) calculations are in
agreement with much of this, but indicate that the electron-phonon
interaction has
insufficient strength $\left( \lambda _s\sim 1\right) $ 
\cite{PRB90,Pickett,Kanazawa,SantaFe} and leads to $s$%
-wave pairing. The experimental evidence that the symmetry of the paired
state is $d$,\cite{Levi} 
with lobes in the direction of the Cu-O bond, points to the
Coulomb repulsion between two holes on the same copper site as the pairing
agent. However, the scarcity of high-temperature superconducting materials,
as well as results of Hubbard- and $t$-$J$-model 
calculations\cite{ScalapinoReview}, lead to the
suspicion that something more than the Coulomb repulsion is needed.

\begin{figure}[]
\unitlength1cm
\begin{minipage}[]{3.2in}
\centerline{
\rotatebox{-90}{\resizebox{3in}{!}{\includegraphics{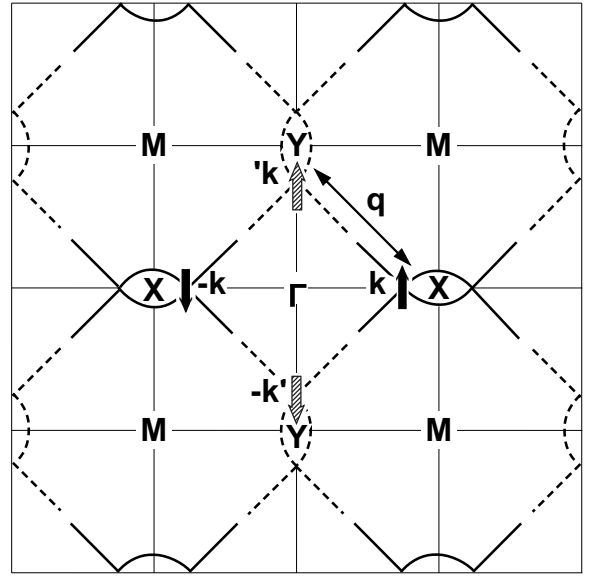}}}
}
\vspace{4 mm} 
\caption{\label{fig1}
{\footnotesize
Fermi surface of an optimally doped CuO$_2$ superconductor
(schematic). The energy bands are nearly flat around X and Y (slightly
bifurcated saddle-points). For the experimentally observed $d_{x^2-y^2}$%
-wave pairing, the superconducting gap $\Delta \left( \mathbf{k}\right) $
has opposite signs on the parts of the Fermi surface which are shown in
full- and broken-line. The scattering of a Cooper pair at $\left( \mathbf{%
k,-k}\right) $ to one at $\left( \mathbf{k}^{\prime },-\mathbf{k}^{\prime
}\right) $ is indicated. The pair-interaction $V\left( \mathbf{k,k}^{\prime
}\right) $ which sustains this gap-anisotropy must be repulsive for momentum
transfers $\mathbf{q\equiv k^{\prime }-k}$ connecting full and broken parts,
and attractive for $\mathbf{q}$'s connecting full with full, and broken with
broken parts. The first property is presumably provided by the Coulomb
repulsion and the second, we argue, could be provided by the interaction
between near-saddle-point electrons and buckling modes.
}}
\end{minipage}
\hfill
\end{figure}

It is therefore of interest\cite{Scalapino,Bulut} 
to investigate whether the electron-phonon interaction,
which gives a negative (attractive) pair-interaction, 
$V_{ep}\left( {\bf k,k}%
^{\prime }\right) \propto -\left| g\left( {\bf k,k}^{\prime }\right) \right|
^2,$ could support the observed singlet- and $d$-wave 
pairing with the gap-anisotropy: $\Delta \left( {\bf %
k}\right) \propto \cos \left( ak_x\right) -\cos \left( ak_y\right) $
sketched in Fig. \ref{fig1}. 
For this to occur, and assuming a BCS-like zero-temperature gap-equation: 
\[
2\Delta \left( \mathbf{k}^{\prime }\right) =-\sum_{\mathbf{k}}V\left( 
\mathbf{k}^{\prime },\mathbf{k;\,}\omega \right) \frac{\Delta \left( \mathbf{%
k}\right) }{\sqrt{\left[ \varepsilon \left( \mathbf{k}\right) -\varepsilon
_F\right] ^2+\Delta \left( \mathbf{k}\right) ^2}}\,, 
\]
the
electron-phonon interaction must be large when 
$\Delta \left( {\bf k}\right) $ and $%
\Delta \left( {\bf k}^{\prime }\right) $ have the same sign, and small when
they have opposite signs. 
The main pairing interaction
is usually believed to be associated with the exchange of spin fluctuations,
because it is repulsive and seems to peak for large momentum transfers \cite
{ScalapinoReview}. In the RPA, its form is
\[
V_{sf}\left( \mathbf{q,}\omega \right) =\frac 32\bar{U}^2\frac{\chi _0\left(
\mathbf{q,}\omega \right) }{1-\bar{U}\chi _0\left( \mathbf{q,}\omega \right)
}
\]
where the band susceptibility $\chi _0\left( \mathbf{q,}\omega \right) $ is
supposed to be peaked near $\mathbf{q}=\left( \frac \pi a,\frac \pi a\right)
.$
On the basis of models derived from LDA
calculations\cite{PRB94,Stanford,Karlsruhe}, 
as well as LDA linear-response calculations\cite{Savrasov}, 
we shall argue
that this is the case for the so-called buckling mode, which for YBa$_2$Cu$_3
$O$_7$ and ${\bf q}${\bf =}0 is the 330 cm$^{-1}$ ($\hbar \omega \sim $40
meV) oxygen out-of-plane and out-of-phase mode. Although the interaction of
electrons with this mode is proportional to the static dimple (YBa$_2$Cu$_3$O%
$_7;$ $\delta =7^{\circ })$ or buckle (doped La$_2$CuO$_4$ LTO phase), and
hence expected to be small, Raman and neutron-scattering experiments, as
well as LDA calculations, have shown it to be relatively strong$.$

Fig. 2 specifies our generic tight-binding model for the 
electronic band structure
of a single CuO$_2$-plane. In the upper part of the 2nd row are shown the
four $\sigma $-orbitals, $\left| y\right\rangle \equiv $ O3$_y,$ $\left|
d\right\rangle \equiv $ Cu$_{x^2-y^2},$ $\left| x\right\rangle \equiv $ O2$%
_x,$ and $\left| s\right\rangle \equiv $ Cu$_{s-\left( 3z^2-1\right) },$ and
in the lower part, the $\left| d\right\rangle $ orbital and two of the $\pi $%
-orbitals, $\left| z\right\rangle \equiv $ O2$_z$ and $\left|
xz\right\rangle \equiv $ Cu$_{xz},$ seen from the side of the plane. When
merely the $pd\sigma $ and $pd\pi $ hopping integrals $t_{xd}$=$t_{yd}\equiv
t_{pd}$=1.6 eV and $t_{z,xz}$=$t_{z,yz}$=0.7 eV indicated in the 1st column
are taken into account, the band structure and corresponding constant-energy
contours, in Fig. 3, shown in the 1st columns arise: The $\sigma $-orbitals give rise to
a bonding, a non-bonding, and an anti-bonding O$_x$--Cu$_{x^2-y^2}$--O$_y$
band plus a high-lying Cu$_{s-\left( 3z^2-1\right) }$ level (full lines),
and the $\pi $-orbitals give rise to two decoupled pairs of bonding
anti-bonding bands which disperse in either the $k_x$ or $k_y$ direction
(stippled lines). 
The orbital energies (in eV and with respect to the energy
of the Cu$_{x^2-y^2}$ orbital) are indicated at the relevant points of the
band structure. The conduction band is the anti-bonding $pd\sigma $ band and
the Fermi level will be close to the saddle-point at X $\left( \pi ,0\right) 
$. In the 1st column, this saddle-point is well above the top of the $pd\pi $%
-band and is {\em symmetric,} in the sense that the absolute values of the
band masses in the $k_x$ and $k_y$ directions are equal. This means that the
constant-energy contour through the saddle-point is a square with corners at
X and Y. This is shown in 1st row of Fig. 3. In the 2nd columns, the
strong hoppings ($t_{sp}$=2.3 eV) between the high-energy Cu$_{s-\left(
3z^2-1\right) }$ orbital and the O2$_x$ and O3$_y$ orbitals 
and the tiny O2$_z$ and O3$_z$ hopping are 
included.
\end{multicols}
\begin{figure}[bt]
\unitlength1cm
\begin{minipage}[t]{6.2in}
\centerline{
\rotatebox{-90}{\resizebox{4.5in}{!}{\includegraphics{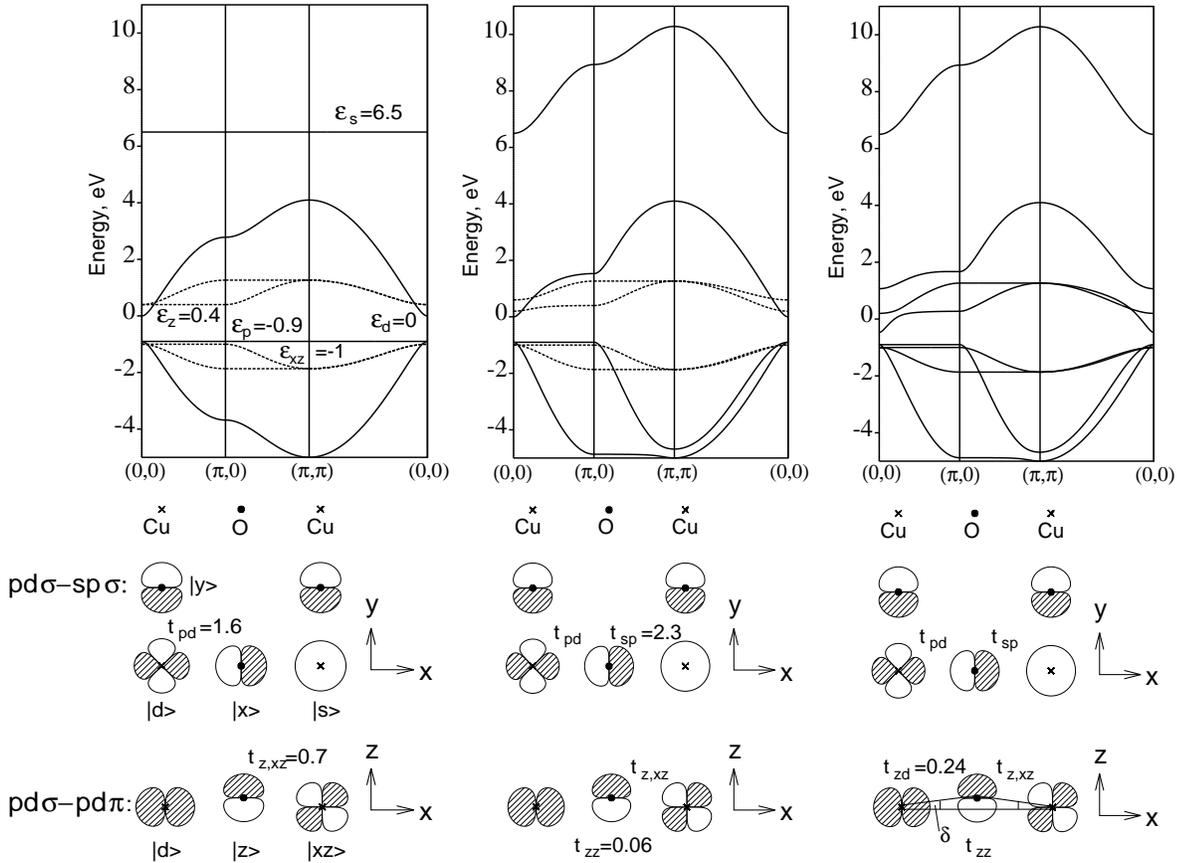}}}
}
\vspace{4 mm}
\caption[]{\label{fig2}
{\footnotesize
Eight-band Hamiltonian for a single CuO$_2$ plane (2nd row) and
synthesis of its band structure (1st row).
The 1st column shows uncoupled $pd\sigma $ (O$_x$%
-Cu$_{x^2-y^2}$-O$_y),$ $sp\sigma $ $\left( \text{Cu}_s\right) ,$ and
$pd\pi
$ (Cu$_{xz}$-O$_z$ and Cu$_{yz}$-O$_z$) bands. In the 2nd column, the
coupling $(t_{sp})$ of the Cu$_s$ orbital to the $pd\sigma $ band is
included. In the 3rd column, also the coupling $\left( t_{zd}\right) $
between $\sigma $- and $\pi $-bands induced by a static dimple $\left(
\delta =7^{\circ }\right) $ is included. Modulation of the latter gives
rise
to interaction with the buckling mode. All energies are in eV.
}}
\end{minipage}
\end{figure}

\begin{multicols}{2}
This depresses the conduction band near $\left( \pi ,0\right) $ and thereby
increases the mass towards $\Gamma $ $\left( 0,0\right) $ and decreases it
towards M $\left( \pi ,\pi \right) ;$ the saddle-point becomes {\em 
asymmetric }and the constant-energy contours now bulge towards $\Gamma .$
With the flat part of the conduction band just straddling off the top of a $%
\pi $-band, even a weak dimple or buckle of the plane will introduce
considerable hybridization between the $\sigma $ and $pd\pi $ bands. This is
seen in the 3rd columns where we have turned on the weak Cu$_{x^2-y^2}$--O$_z$
hoppings ($t_{zd}$=0.24 eV $\propto \sin \left( \delta \text{=}7^{\circ
}\right) $). Since the hybridization with the Cu$_{xz}$--O2$_z$ $pd\pi $
band (but not with the lower-lying Cu$_{yz}$--O3$_z$ band) vanishes at X $%
\left( \pi ,0\right) $, the saddle-point {\em bifurcates} away from X and
towards $\Gamma ,$ once $\delta $ exceeds a critical value ($\sim 4^{\circ }$
in YBa$_2$Cu$_3$O$_7).$ The conduction band thereby becomes very flat in a
region around X (and Y) extending towards $\Gamma .$
\end{multicols}
\begin{figure}[bt]
\unitlength1cm
\begin{minipage}[t]{6.2in}
\centerline{
\rotatebox{-90}{\resizebox{4.0in}{!}{\includegraphics{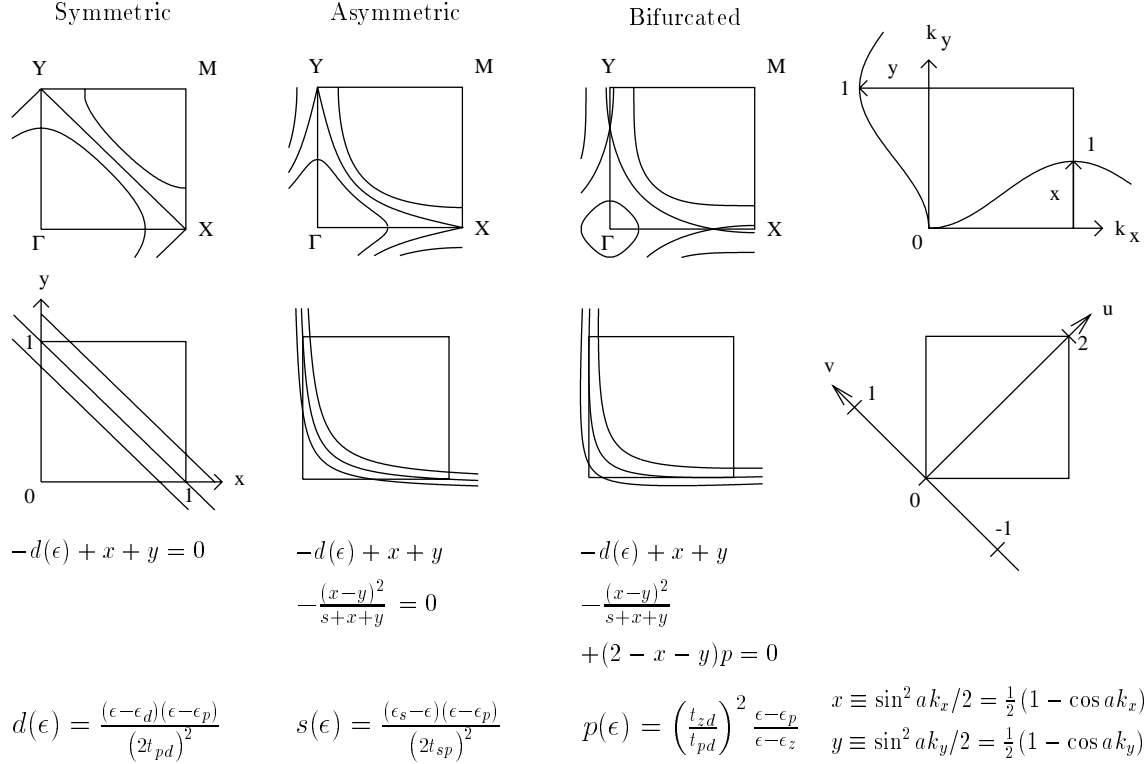}}}
}
\vspace{4 mm}
\caption[]{\label{fig3}
{\footnotesize
Constant-energy contours in $
\left( k_x,k_y\right) $-space (1st row), as well as in an $\left( x,y\right) 
$-space where the constant-energy contours are hyperbolas (2nd row). The
constant-energy contours shown are those of the second band from the 
top in Fig. 2 and
for energies close to that of the saddle-point at X $\left( \pi ,0\right) ,$
which is the Fermi level for optimal doping. The coordinate transformation
is specified in the 4th column and the analytical expressions for the
constant-energy contours are given in the 3rd and 4th rows. The functions $%
d\left( \epsilon \right) ,\;s\left( \epsilon \right) ,$ and $p\left(
\epsilon \right) $ describe respectively the $pd\sigma ,$ $sp\sigma ,$ and $%
\sigma \pi $ scatterings.
In the first column $s$=$\infty $, and $p$%
=0. In the second column $s$=0.6 and $p$=0. In the third column, $s$=0.6
and $ p $=0.14.
}}
\end{minipage}
\end{figure}

\begin{multicols}{2}

It is this near degeneracy with the top of a $\pi $-band of the saddle-point
of the conduction-band, which has $\sigma $-character, which causes strong
coupling to out-of-plane deformations. How near the $\sigma $-$\pi $
degeneracy is, is first of all controlled by the energy of the Cu$_{s-\left(
3z^2-1\right) }$ hybrid and, hence, by the distance of apical oxygen from
plane copper. Secondly, if the saddle-point of the $\sigma $-band is
suppressed below the top of the $\pi $-band and part of the Fermi-surface
becomes $\pi $-like, then a static dimple or buckle will adjust itself
through reduction of the $\pi $-band width $\left( \propto t_{z,zx}\right) $
so that the $\pi $-band becomes essentially filled. This, we have learned
from total-energy LDA calculations\cite{Savrasov}.

Although our tight-binding model has 8 orbitals per CuO$_2$ unit, the
constant-energy contours, $\epsilon _j\left( k_x,k_y\right) =\epsilon ,$
have simple analytical expressions. These are given in the bottom rows of
Fig. 3 and are expressed in terms of functions, $d\left( \epsilon \right)
,\;s\left( \epsilon \right) ,$ and $p\left( \epsilon \right) ,$ which
describe the $dp\sigma ,$ $sp\sigma ,$ and $\sigma \pi $ scatterings. If we
neglect the Cu$_{xz}$ and Cu$_{yz}$ orbitals the constant-energy contours
are merely hyperbolas in a $\left( x,y\right) $-space defined by: $x\equiv 
\frac 12\left( 1-\cos ak_x\right) \;$and $y\equiv \frac 12\left( 1-\cos
ak_y\right) ,$ and they are shown in the 2nd row of Fig. 3. The bottom rows
give the explicit expressions. With this model we can now easily calculate 
\[
\begin{array}{lll}
\chi _0^{\prime \prime }\left( {\bf q,}\omega \right)& =&\,2\pi \int \frac{d^2k%
}{BZA}\left[ f\left( \epsilon \left( {\bf k}\right) \right) -f\left(
\epsilon \left( {\bf k+q}\right) \right) \right] \\
\\
&&\times \delta \left( \epsilon
\left( {\bf k+q}\right) -\epsilon \left( {\bf k}\right) -\hbar \omega
\right) 
\end{array}
\]
as a function of the band shape and the doping. For ${\bf q\sim
Q\equiv }\left( \pi ,\pi \right) ,$ where the Coulomb repulsion is supposed
to provide the pairing interaction, $\chi _0^{\prime \prime }$ vanishes when 
$\epsilon _F$ is below the saddle-point and $\hbar \omega $ is less than $%
\sim \epsilon _{{\rm saddle}}-\epsilon _F$ (overdoping). If $\epsilon _F$ is
at or above a saddle-point and this is slightly bifurcated, then $\chi
_0^{\prime \prime }\left( {\bf q,}\omega \right) $ has a large, broad peak
around ${\bf Q}$ for $\hbar \omega $ larger than $\sim \epsilon _F-\epsilon
_{{\rm saddle}}.$ This peak is due to ''nesting'' of the flat band near X
with the one near Y. This, we believe, gives rise to spin-fluctuations which
provide the repulsive, large-${\bf q}${\bf \ }part of the pairing
interaction. For smaller ${\bf q}$ and near-optimal doping, $\chi _0^{\prime
\prime }\left( {\bf q,}\omega \right) $ with $\hbar \omega $ $\sim $40 meV
has high ridges which are caused by the ''sliding'' of a saddle-point along
the Fermi surface. We shall now show that the bucling phonon takes advantage
of this part of $\chi _0^{\prime \prime }$ and does not feel the peak at
large ${\bf q}$ because here, its interaction $\left| g\left( {\bf k,k}%
^{\prime }\right) \right| ^2$ vanishes.

The last point is illustrated in Fig. 4 where we have assumed a static dimple
like in YBa$_2$Cu$_3$O$_7{\bf .}$ The left- and right-hand sides show the
electronic wave-functions at X and Y, respectively. The Cu$_{s-\left(
3z^2-1\right) }$ character, irrelevant for our qualitative argument, is
neglected, but the weak Cu$_{yz}$-O3$_z$ and Cu$_{xz}$-O2$_z$ characters at
respectively X and Y are included. As mentioned above, and as can be seen
from the figure, there can be no Cu$_{xz}$-O2$_z$ character at X and no Cu$%
_{yz}$-O3$_z$ character at Y. On the right-hand side, we have perturbed the
wave-function at Y by the buckling mode with ${\bf q=Q.}$ This wave in the $%
\left[ 1,1\right] $-direction increases/decreases the dimple of every other
oxygen row and accordingly modulates the coupling $t_{zd}$ between each O2$_z
$ and its neighboring Cu$_{x^{2-}y^2}$ orbitals. The electron-phonon matrix
element $\left\langle \Psi \left( {\rm X}\right) \right| \delta \left( {\bf Q%
}\right) \left| \Psi \left( {\rm Y}\right) \right\rangle $ is seen to have
contributions from hopping between orbitals,
drawn with thick lines, only and it vanishes because 
$\delta \left( {\bf Q}\right) \left| \Psi \left( {\rm Y}\right)
\right\rangle $ is even and $\left\langle \Psi \left( {\rm X}\right) \right| 
$ is odd with respect to a $y$-axis through the O2's.
\begin{figure}[bt]
\unitlength1cm
\begin{minipage}[t]{3.2in}
\centerline{
\rotatebox{0}{\resizebox{3in}{!}{\includegraphics{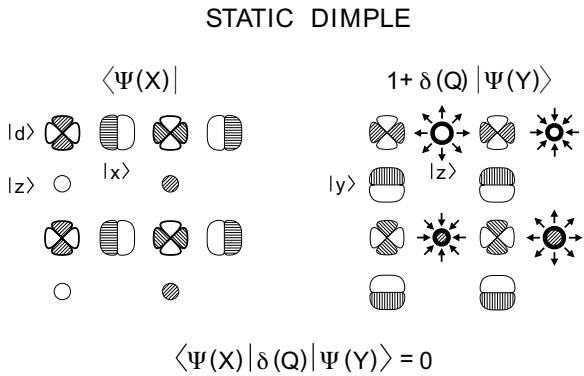}}}
}
\vspace{4 mm}
\caption[]{\label{fig4}
{\footnotesize
Schematic illustration of why there is no electron-phonon
interaction for the
bucling mode with ${\bf Q\equiv }\left( \pi ,\pi \right) $.
}}
\end{minipage}
\end{figure}

The strength of the electron-phonon interaction is 
\[
\lambda =\left[ \pi N\left(
\epsilon _F\right) \right] ^{-1}\int_{BZ}\frac{d^2q}{BZA}\gamma \left( {\bf %
q,}\omega \right) /\omega \left( {\bf q}\right) ^2
\]
where 
\[
N\left( \epsilon
_F\right) =\int_{FS}\frac{dk}{BZA}\left| v\left( {\bf k}\right) \right| ^{-1}
\] 
is the electronic density of states and
\[
\begin{array}{lll}
\gamma \left( {\bf q,}\omega \right)& =&2\pi \int \frac{d^2k}{BZA}\left\{
f\left[ \epsilon \left( {\bf k}\right) \right] -f\left[ \epsilon \left( {\bf %
k+q}\right) \right] \right\}  \\
\\          
&&\times \delta \left[ \epsilon \left( {\bf k+q}\right)
-\epsilon \left( {\bf k}\right) -\hbar \omega \left( {\bf q}\right) \right]
\left| g\left( {\bf k,k+q}\right) \right| ^2 
\end{array}
\]
the phonon linewidth. For the $d$-wave channel with gap-anisotropy $\frac 12%
\left( \cos ak_y-\cos ak_x\right) =x-y,$ a factor $\left( x-y\right) ^2$
must be included in the integrand for $N\left( \epsilon _F\right) ,$ thus
yielding $N_d\left( \epsilon _F\right) ,$ and a factor $\left( x-y\right)
\left( x^{\prime }-y^{\prime }\right) $ must be included in the integrand
for $\gamma \left( {\bf q,}\omega \right) .$ Hence for $g$ constant, $%
\lambda _d$ would vanish. For the buckling mode our tight-binding model
yields:
\[
\begin{array}{lll}
g\left( {\bf k,k}^{\prime }\right)& =&2t_{zd}\,\frac{\epsilon _F-\epsilon _p}{%
\left( \epsilon _F-\epsilon _z\right) \left( \epsilon _F-\frac{\epsilon
_p+\epsilon _d}2\right) }\,\frac{\partial t_{zd}/\partial z_{\text{O}}}{%
\sqrt{M\omega }}\;     \\
\\         
&&\times \left[\sqrt{\left( 1-x\right) \left( 1-x^{\prime
}\right) }-\sqrt{\left( 1-y\right) \left( 1-y^{\prime }\right) }\right] ,
\end{array}
\]
where for simplicity we have neglected the dispersion of the $\pi $ bands.
It is easy to see that this form vanishes for ${\bf q=Q,}$ for instance. To
further illustrate this form we 
\begin{figure}[bt]
\unitlength1cm
\begin{minipage}[t]{3.2in}
\centerline{
\rotatebox{-90}{\resizebox{3in}{!}{\includegraphics{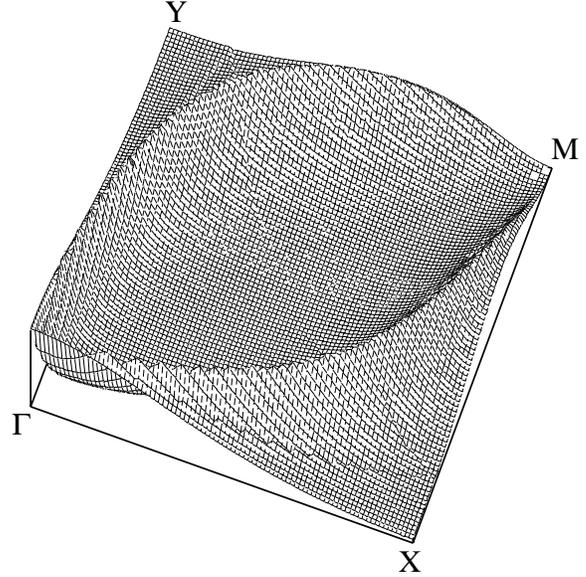}}}
}
\vspace{4 mm}
\caption[]{\label{fig5}
{\footnotesize
$\left( \mathbf{k-k}^{\prime }\right) $-dependence of the
electron-phonon interaction for the out-of phase buckling mode from the
six-orbital model. $\mathbf{k}$ and $\mathbf{k}^{\prime }$ were chosen on
the YBa$_2$Cu$_3$O$_7$-type Fermi surface,
which lies between the two low
hole doping constant-energy contours in the third column of
Figure 3.
}}
\end{minipage}
\end{figure}

\noindent
show in Fig. 5 
$g\left( {\bf q}\right) \equiv
\gamma \left( {\bf q},0\right) /\chi _0^{\prime \prime }\left( {\bf q,}%
0\right) ,$ calculated for a band shape with a marginally bifurcated
(extended) saddle-point slightly below $\epsilon _F.$ This $g$ is seen to
have the kind of dispersion required in order to support $d$-wave pairing. $%
g\left( {\bf k,k}^{\prime }\right) $ as given above, factorizes and the
double-integral for $\lambda _d$ may be expressed analytically and the
result for $\omega \rightarrow 0$ is simply: $\lambda _d=\,c^2\omega
^{-1}N_d\left( \epsilon _F\right) ,$ where $c$ is the factor in front of the
square parenthesis in the expression for $g\left( {\bf k,k}^{\prime }\right)
.$ This result is {\em positive definite} for all dopings. For $\epsilon _F$
approaching a saddle-point at X, $N_d\rightarrow N,$ so that the strength of
the electron-phonon interaction 
in this $d$-channel is as large as in the $s$-channel.

Finally, in Fig. 6 we show the phonon spectrum calculated {\em ab initio}
with the linear-response LDA-LMTO method for the idealized
infinite-layer compound CaCuO$_2$ doped with 0.24 holes per 
formula unit\cite{Savrasov}. In
the calculations, this compound developed a static buckle (O2 up and O3 down
by $\delta $=6$^{\circ }$). The uppermost of the three phonon branches
marked with spheres is the buckling mode and the numbers written on the
modes are the mode $\lambda _d$'s. We see that the buckling mode dominates $%
\lambda _d$ and that it is large close to $\Gamma $ along $\Gamma $X and
vanishes at M $\left( \pi ,\pi \right) $. Numerically we found $\lambda
_d=0.3$ and $\lambda _s=0.4.$
\begin{figure}[bt]
\unitlength1cm
\begin{minipage}[t]{3.2in}
\centerline{
\rotatebox{0}{\resizebox{3in}{!}{\includegraphics{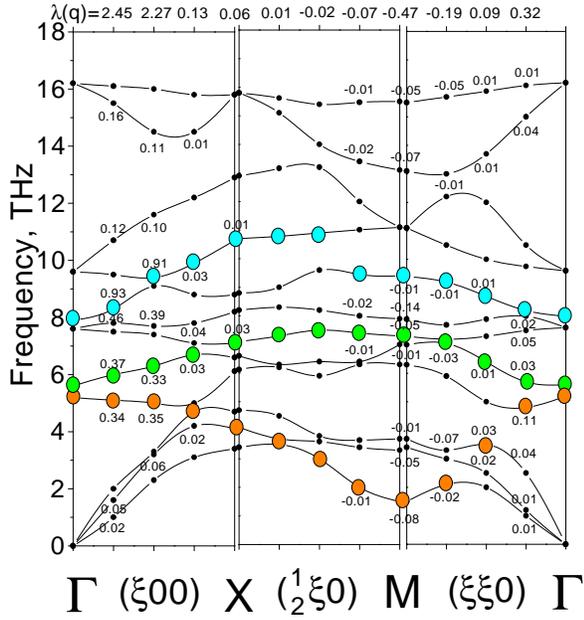}}}
}
\vspace{4 mm}
\caption[]{\label{fig6}
{\footnotesize
Phonon modes and the mode $\lambda _d$ calculated {\it ab
initio}
with the linear-response LDA-LMTO method for CaCuO$_2$ doped
with 0.24 holes. 
}}
\end{minipage}
\end{figure}

We conclude that the linear electron-phonon interaction for buckled or
dimpled CuO$_2$-planes may support, but is hardly sufficient to cause,
high-temperature superconductivity based on $d_{x^2-y^2}$-pairing. The most
important mode seems to be the buckling mode, because it modulates the
saddle-points of the energy bands where the density of states is high and
where the superconducting gap is observed to be maximum.
The electron-phonon interaction, for this mode,
is small or vanishes for momentum
transfers {\bf q} near ($\pi $,$\pi $), and does therefore not interfere
with the $d$--wave symmetry of the order parameter induced by 
spin-fluctuation exchange.

\end{multicols}

\end{document}